\documentclass[twocolumn,showpacs,preprintnumbers,amsmath,amssymb]{revtex4}

\usepackage{graphicx}
\usepackage{dcolumn}
\usepackage{bm}

\def \be{\begin{equation}}
\def \ee{\end{equation}}
\def \bea{\begin{eqnarray}}
\def \eea{\end{eqnarray}}
\def \ba{\begin{array}}
\def \ea{\end{array}}
\def \non{\nonumber}

\def \ben{\begin{enumerate}}
\def \een{\end{enumerate}}

\begin{document}


\title{The Role of Friction in Compaction and Segregation of Granular Materials}
\author{Yair Srebro and Dov Levine}
\affiliation{Department of Physics, Technion, Haifa 32000, Israel}

\date{\today}

\begin{abstract}

We investigate the role of friction in compaction and segregation
of granular materials by combining Edwards' thermodynamic
hypothesis with a simple mechanical model and mean-field based
geometrical calculations. Systems of single species with large
friction coefficients are found to compact less. Binary mixtures
of grains differing in frictional properties are found to
segregate at high compactivities, in contrary to granular mixtures
differing in size, which segregate at low compactivities. A phase
diagram for segregation vs. friction coefficients of the two
species is generated. Finally, the characteristics of segregation
are related directly to the volume fraction without the explicit
use of the yet unclear notion of compactivity.

\end{abstract}

\maketitle


\section{Introduction:}

Friction plays a significant role in packings of grains in a
granular material, however due to the non-equilibrium nature of
such materials their macroscopic properties do not trivially
result from the microscopic physics of their constitutes.
Compaction and segregation are two macroscopic phenomena which
occur in granular materials undergoing series of extensive
operations \cite{edwards_oakeshott_physica_a_1989} (operations of
a statistical nature rather than ``Maxwell-demon'' operations
which act on single grains). Packings of identical hard spheres
compacted by extensive operations reach a state of random close
packing (RCP), rather than an ordered crystalline packing, which
has a higher volume fraction (defined as the ratio of the sum of
grain volumes to the volume the system occupies). Understanding
the process leading to RCP, its geometrical properties and whether
higher volume fractions may be obtained is of great importance for
physics and for engineering. The tendency towards segregation in
granular mixtures comprised of grains with various mechanical
properties is interesting for the physicist and disturbing for the
engineer, for whom a homogeneous mixture is often an industrial
need (for a review see \cite{kaye}).

Segregation in granular materials has received much attention in
recent years (for a review see \cite{ristow}). The phenomenon is
observed for grains varying in size \cite{cantelaube_1995}, shape
\cite{lawrence_1968}, friction coefficient \cite{zik_1994} and
density \cite{shinbrot_1998}. Segregation occurs due to vibration
\cite{cooke_1996}, tapping \cite{knight_1993}, rotation
\cite{zik_1994}, pouring \cite{makse_nature_1997} and shearing
\cite{williams_1976}. Experiments are performed for mixtures of
many particles of two different species (for a review see
\cite{ottino_2000}) and with single intruder particles in systems
of a single species \cite{clement_1995}. Existing theoretical
modeling of segregation due to rotation
\cite{zik_1994,levine_chaos_1999,lai_1997} and to pouring
\cite{makse_nature_1997,boutreux_de_gennes_1996} is based on
kinetic phenomena: segregation is explained as a result of
different flow properties of the different species. Monte Carlo
simulations of dynamic phenomena in vibrated systems give insight
into segregation \cite{rosato_1987} as well as compaction dynamics
of single species systems \cite{mehta_barker_1991}. Similar
dynamic phenomena have been captured analytically in models based
on free volume considerations
\cite{boutreux_de_gennes_1997,ben_naim_1998,nowak_1998}.

This paper deals with the role of friction in compaction and
segregation through the analysis of the static properties of
granular materials. The results obtained may be used in order to
verify experimentally the validity of such models. One such
proposal for the description of static granular materials is the
analogy to the statistical mechanics of thermodynamic of Edwards
\cite{edwards_oakeshott_physica_a_1989}, a more detailed
description of which will be given in the following section. The
central idea behind it is that even though the system is static
and does not move with time within the ensemble of mechanically
stable arrangement of the grains, we may assume ergodicity and
employ statistical mechanics considerations for the probability of
finding the system in any one of its states. This model requires
the existence of an analog of temperature, referred to as
``compactivity'' (Other effective temperatures may be defined for
jammed granular materials \cite{langer_liu_2000,ono_liu_2002}, and
attempts have been done to connect them to the compactivity
\cite{barrat_2000,makse_nature_2002}). Recent experimental
evidence for reversibility in compaction processes
\cite{nowak_1997} has provided justification for this
thermodynamic analogy and have proposed a connection between
compactivity and experimentally controllable quantities
\cite{edwards_grinev_PRE_1998}, however the proposal remains
controversial.

Edwards' hypothesis is that an analog of the free energy $Y=V-XS$
is minimized where $V$ is the system volume, $S$ is the entropy
and $X$ is the compactivity. Our expectation that this formalism
may predict frictional segregation is based on the following
argument. Consider two systems of identical grains, such that the
friction coefficient of the first system, $\mu_1$, is greater than
that of the second system, $\mu_2$. Then $S_1>S_2$ because every
configuration available to the second system may be identically
realized to the first system, while the converse is not true. This
suggests that under certain circumstances a mixture of grains with
different friction coefficients may prefer to segregate in order
that its entropy be maximized, competing with the preference of
the entropy of mixing to be maximized in the homogeneously mixed
state.

The statistical hypothesis has been used to investigate
segregation in binary mixtures of species differing in size by
mapping them to the Ising model, resulting in segregation below
some critical compactivity
\cite{edwards_oakeshott_physica_a_1989,mehta_edwards_physica_a_1989}.
Recent geometrical calculations have enabled relating the ideas of
the statistical proposal to actual densities of granular systems
\cite{blumenfeld_2003}.

In section \ref{sec:one_spec} we use Edwards' statistical
hypothesis together with a simple mechanical model for the
quantitative description of friction in 2D and 3D single species
granular materials. This is combined with simple calculations of
Voronoi cell volumes for general coordination numbers resulting in
the dependence of volume fraction on friction coefficient and
compactivity. Section \ref{sec:two_spec} uses the mean-field
approximation in order to describe segregation in a binary mixture
of grains differing in frictional properties. Unlike mixtures of
grains differing in size, which may be mapped to the Ising model
\cite{edwards_oakeshott_physica_a_1989,mehta_edwards_physica_a_1989},
frictional differences between grains result in larger entropy for
the rougher grains. Therefore, these systems may not be mapped
exactly onto the Ising model and segregation occurs above a
critical compactivity and not below it. We then generate a phase
diagram for segregation vs. friction coefficients of the two
species. Finally, the dependence of the results on compactivity is
eliminated by using the volume fraction as a measure for
compactivity. Section \ref{sec:conc} concludes with a summary of
the results.


\section{Single Species:}
\label{sec:one_spec}

\subsection{Statistical Model}

We use the statistical mechanics hypothesis proposed by Edwards
\cite{edwards_oakeshott_physica_a_1989} for the description of
jammed granular systems. In this formalism each mechanically
stable arrangement of the grains is equivalent to a micro-state in
statistical mechanics, and the total volume the grains occupy
plays the role of the energy. The analog of temperature is assumed
to exist, is denoted X and is called the compactivity. We measure
X in units of volume so that the analog of the Boltzman constant
is equal to unity. In analogy with the canonical ensemble of
states in thermal systems, the probability for the occurrence of a
state with volume V is assumed to be proportional to $ e^{-V/X} $.

We consider a system of N identical spherical (in 3D) or circular
(in 2D) grains. For every arrangement of the grains, the total
volume of the system may be written as the sum over all grains of
the Voronoi cell volume around each grain, $ v_i $: \bea
V=\sum^{N}_{i=1} v_i . \eea

Average volumes of Voronoi cells, calculated neglecting spatial
correlations between locations of grains, have been shown to agree
with exact calculations \cite{richard_2001}. Moreover,
correlations have been shown to have a small effect on the
dependence of total volume on compactivity \cite{blumenfeld_2003}.
Therefore, we use a mean-field approximation and assume the volume
of every Voronoi cell is evenly distributed between a minimal and
maximal volume, $ v_{min} $ and $ v_{max} $. The geometrical and
mechanical considerations determining these volumes will be
presented in the following section, and at this stage it is only
assumed $ v_{min} $ and $ v_{max} $ are uniform for all grains in
the system.

Following the analogy with the canonical ensemble, we may
calculate the pseudo partition function: \bea Z &=&
\int^{v_{max}}_{v_{min}} \cdot \cdot \cdot \int
^{v_{max}}_{v_{min}} e^{-\sum v_i
/ X} dv_1 \cdot \cdot \cdot dv_N = \non \\
&=& \left( \int ^{v_{max}}_{v_{min}} e^{-v/X} dv \right)^N = \non \\
&=& \left(2X \cdot e^{-v_{mid}/X} \cdot \sinh\left( \frac {\Delta
v}{X}\right)\right)^N \label{eq:part_fun_one_specie}, \eea where
we have introduced the notations, $
v_{mid}\equiv (v_{min}+v_{max})/2 $ and $ \Delta v
\equiv (v_{max}-v_{min})/2 $.

The average volume per grain may easily be derived from the
partition function as: \bea <v> &=& \frac {<V>}{N} = \frac {1}{N}
X^2 \frac {\partial\ln(Z)}{\partial X} = \non \\
&=& v_{mid} + X - \Delta v \cdot \coth \left(\frac{\Delta
v}{X}\right)\label{eq:avrg_vol_one_specie}. \eea This expression
has been derived in \cite{edwards_oakeshott_physica_a_1989}, and
it is clearly seen that as $X \rightarrow 0$, $<v> \rightarrow
v_{min}$ and as $X \rightarrow \infty$, $<v> \rightarrow v_{mid}$.
This is analogous to thermal systems, where at low temperatures
the system is most probable to be found in its ground state, and
as the temperature is increased exited states are occupied with
increasing probability until the limit of infinite temperature,
where all states are occupied with an equal probability, and the
system's energy is the average energy of all these states.


\subsection{Mechanical Model}

We would now like to introduce mechanical and geometrical
considerations to estimate $v_{min}$ and $v_{max}$, which must be
known in order to evaluate the expression in eq.
(\ref{eq:avrg_vol_one_specie}). The minimal volume is achieved for
hexagonal packing in 2D and face centered cubic or hexagonal close
packing in 3D. The corresponding Voronoi cell volumes are $
v_{min}^{2D}=\sqrt{12} r^2 $ and $ v_{min}^{3D}=\sqrt{32} r^3$,
where r is the grain radius. The resulting volume fractions are $
\Phi_{max}^{2D} = {\pi r^2}/{v_{min}^{2D}} = 
{\pi}/{\sqrt{12}} \simeq 0.91$ and $\Phi_{max}^{3D}= {\frac {4
\pi}{3} r^3}/{v_{min}^{3D}} = {\pi}/{\sqrt{18}} \simeq 0.74$.

Although purely geometric considerations determine $v_{min}$, the
frictional forces between the grains manifest themselves in $
v_{max} $. The idea behind this is that friction at grain contacts
allows for the formation of arcs and for a gradual decrease in the
number of contacts per grain \cite{silbert_2002}, which in turn
increases the volume of the Voronoi cell around every grain.

In order to estimate the effect of friction in granular materials
we will consider ``toy-systems'' consisting of a small number of
grains (three in 2D and four in 3D), calculate the effect of
friction there, and use the result in order to obtain an
approximate prediction for the dependence of $ v_{max} $ on the
friction coefficient, $\mu$. This, in turn, may be inserted
together with $v_{min}$ into eq. (\ref{eq:avrg_vol_one_specie}) in
order to obtain an approximate expression for the dependence of
the total volume of the system, and hence of the volume fraction,
on $\mu$. The results will depend on the compactivity, whose
physical significance still requires elucidation. However, even
without understanding its significance, a few predictions for
experimental results may be drawn from the model presented here.

\begin{figure}[htb]
\includegraphics[width=8cm]{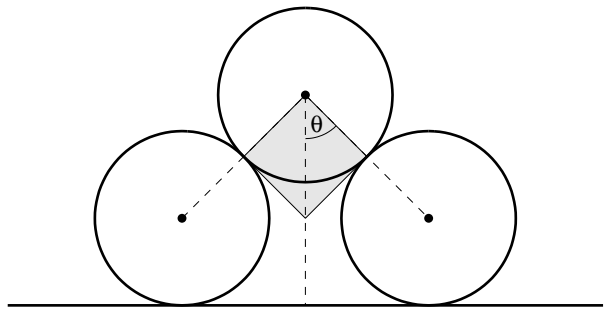}
\caption{\label{fig:basic_balls_2d} The 2D ``toy-system''. The
shaded rhombus is the segment of the Voronoi cell lying between
two adjacent contacts.}
\end{figure}

The 2D ``toy-system'' consists of two grains lying on top of a
horizontal plane and a third grain lying on top of these (see fig.
\ref{fig:basic_balls_2d}). All grains and the horizontal plane are
assumed to be hard, and frictional forces with an equal
coefficient of friction, $\mu$, act at all four contacts. A
uniform gravitational force acts downwards on all grains. The
condition for mechanical equilibrium is that forces and torques
acting on all grains vanish. It can easily be seen that this is
satisfied whenever: \bea \frac {\sin \theta}{1+ \cos \theta} \leq
\mu \label{eq:tet_mu},\eea where $\theta$ is half the angle
between contacts of the top grain with the bottom grains (see fig.
\ref{fig:basic_balls_2d}). For $\mu \geq 1$ eq. (\ref{eq:tet_mu})
is always satisfied, and any state with $ {\pi}/{6} \leq
\theta \leq {\pi}/{2}$ is mechanically stable. For smaller
values of $\mu$ only states with $\theta \leq \theta_{max} $ are
mechanically stable, where $\theta_{max}$ is determined from \bea
\frac {\sin \theta_{max}}{1+ \cos \theta_{max}} = \mu
\label{eq:tet_max_mu}.\eea If $\mu \leq (2+\sqrt{3})^{-1} \simeq
0.3$, the frictional forces cannot hold the top grain on top of
the two bottom ones, and these simple considerations may not be
used in order to determine $v_{max}$. In this case the volume of
the Voronoi cell around every grain is set in the model to
$v_{min}$. Substituting this into eq.
(\ref{eq:avrg_vol_one_specie}) yields $<v>=v_{min}$, which
corresponds to the maximal volume fraction quoted earlier, and
hence, to crystallization. Therefore, this model may not be used
for frictionless systems, since it predicts crystallization at
every compactivity.

\begin{figure}[htb]
\includegraphics[width=8cm]{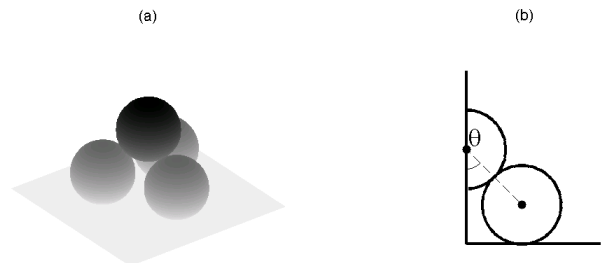}
\caption{\label{fig:basic_balls_3d} a) The 3D ``toy-system''. b)
Section of the top grain and one of the bottom grains.}
\end{figure}

In 3D we consider three grains in an equilateral triangle lying on
top of a horizontal plane and a fourth grain lying on top of them
(see fig. \ref{fig:basic_balls_3d}a). As in the 2D case, all
grains and the horizontal plane are assumed to be hard, frictional
forces with an equal coefficient of friction, $\mu$, act at all
six contacts, and a uniform gravitational force acts downwards on
all grains. Again, the condition for mechanical equilibrium can
easily be seen to be given by eq. (\ref{eq:tet_mu}), however now
$\theta$ is the angle between the vertical direction and the line
connecting the top grain with any one of the bottom grains (see
fig. \ref{fig:basic_balls_3d}b). As in 2D, for small values of
$\mu$, and specifically for frictionless systems, these
considerations cannot be used to estimate the volume fraction,
since this model predicts crystallization regardless of
compactivity, while actual 3D granular systems do not fully
crystallize, but fall into an RCP state.

We would now like to use the results of these simple
considerations to evaluate the maximal volume of the Voronoi cell
around every grain in a granular material. In such systems
frictional forces allow the existence of large angles between
contacts of every grain with its surrounding grains. We assume
that the maximal angle between contacts in a granular material
depends on the friction coefficient in a similar manner to its
dependence in the toy-systems. Moreover, in order to calculate the
Voronoi cell volume around every grain we assume the angles
between adjacent contacts of this grain with its surrounding
grains are uniform and are given by $\theta$ as in the
toy-systems. Note that since we allow $\theta$ to vary
continuously, it may not necessarily be physically possible to
build these packings, even locally. In 2D it is possible only if
$\theta$ is an integer fraction of $\pi$, while in 3D it is
possible only if the coordination number is 4 or 6, which
correspond to $\theta \simeq 1.23$ and $\theta \simeq 0.96$,
respectively. The general expression for the Voronoi cell volume
is (see appendix \ref{sec:geom_calc}): \bea v(\theta)= \left\{
\begin{array}{cc}
  \frac{ \pi r^2 \tan \theta}{\theta} & \mbox{(2D)} \\
  \frac{\pi r^3 \sqrt{3} \sin\theta\tan\theta}
  {(1+3\cos^2\theta)\tan^{-1}
  (\sqrt {\tan \frac{3\alpha}{4} \tan^3 \frac{\alpha}{4}})}
  & \mbox{(3D)}
\end{array} \right.\label{eq:vol_vs_theta},
\eea where $\alpha=\cos^{-1}((3\cos2\theta+1)/4)$ is the
angle between two adjacent contacts (see fig.
\ref{fig:vor_cell_3d}).


\subsection{Results}

\begin{figure}[htb]
\includegraphics[width=8cm]{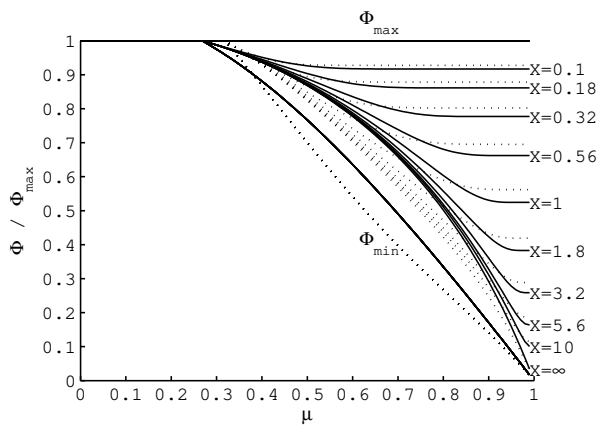}
\caption{\label{fig:vof_vs_x_and_mu} Volume fraction, $\Phi$, vs.
friction coefficient, $\mu$, for several values of the
compactivity, X, for 2D (solid lines) and for 3D (dotted lines).
Volume fractions are normalized according to their maximal values,
$ \Phi_{max}= {\pi}/{\sqrt{12}} \simeq 0.91$ in 2D and
$\Phi_{max}= {\pi}/{\sqrt{18}} \simeq 0.74$ in 3D. The
bounding minimal and maximal volume fractions, $\Phi_{min}$ and  
$\Phi_{max}$, are also plotted.}
\end{figure}

We combine eq. (\ref{eq:tet_max_mu}) and (\ref{eq:vol_vs_theta})
to get the dependence of $v_{max}$ on $\mu$. The resulting volume
fraction, $\Phi$, is now calculated using eq.
(\ref{eq:avrg_vol_one_specie}) and plotted in fig.
\ref{fig:vof_vs_x_and_mu} as a function of friction coefficient
and compactivity. Compaction depends on friction only in the
region $0.3 \lesssim \mu<1$, where the mechanical model used here
is relevant. As $X \rightarrow 0$, $\Phi$ approaches its maximal
value (determined from geometrical considerations), and as $X
\rightarrow \infty$, it approaches a value larger than its minimal
value (determined from mechanical considerations), since all
possible volumes between the minimal and maximal are equally
probable. Obviously, $\Phi$ decreases as either $\mu$ or $X$
increases. The dependence of volume fraction on friction predicted
here may be investigated experimentally, by comparing granular
packings differing only in friction coefficient, which have
otherwise been prepared identically, and hence may be reasonably
assumed to have equal compactivities.

\section{Two Species: segregation}
\label{sec:two_spec}

\subsection{Mean-Field Model}

We would now like to describe a system consisting of two species
of grains differing only in frictional properties. The central
question we wish to address in such systems is whether the two
species mix homogeneously, or segregate into separate domains. For
a system of two species denoted A and B, we shall assume we know
the compactivity, X, and the friction coefficients, $\mu_{AA}$,
$\mu_{AB}$ and $\mu_{BB}$, between two grains of type A, between a
grain of type A and a grain of type B, and between two grains of
type B, respectively. The considerations presented in the previous
section may be used in order to calculate the maximal volumes,
$v_{AA} \equiv v_{max}(\mu_{AA})$, $v_{AB} \equiv
v_{max}(\mu_{AB})$ and $v_{BB} \equiv v_{max}(\mu_{BB})$, of the
Voronoi cells around an A grain surrounded by A grains, an A grain
surrounded by B grains (or a B grain surrounded by A grains) and a
B grain surrounded by B grains, respectively. We describe a
monodisperse system, hence the corresponding minimal volumes are
identical for all types of grains and are denoted here by
$v_{min}$.

We would now like to write the partition function, Z, for two
species and to derive from it the analog of free energy,
$Y=-X\log(Z)$, as a function of the concentration
$f=\frac{N_A}{N}$, where N is the total number of grains and $N_A$
is the number of A grains. The constraint on the total number of
grains of each species in the system causes us to view $f$ as a
local concentration which varies throughout the system. As in the
mean-field description of a binary alloy, which is equivalent to
the Ising model, a single minimum of $Y(f)$ means the two species
tend to get mixed homogeneously at a concentration equal to the
global concentration determined from the number of grains of each
species. The existence of two minima of $Y(f)$ means that the
system tends to separate into domains with two different
concentrations. The percentage of the system with each of these
two minimizing local concentrations is determined from the global
concentration according to the Maxwell construction (see e.g.
\cite{christian}).

The number of A grains is $N_A=fN$ and the number of B grains is
$N_B=(1-f)N$. In the mean-field approximation the number of A-A
contacts is $N_{AA}=f^2Nz/2$, the number of A-B contacts is
$N_{AB}=f(1-f)Nz$ and the number of B-B contacts is
$N_{BB}=(1-f)^2Nz/2$, where $z$ is the average number of neighbors
per grain, or the average coordination number, which is assumed to
be uniform for all types of grains. The contribution of every
contact to the total volume is limited according to its type
between $2v_{min}/z$ and $2v_{ij}/z$, with i and j denoting A or B
for the types of the two grains in contact. The pseudo partition
function is: \bea &Z& = \frac{N !}{(fN) ! ((1-f)N) !} \left(
\int^{v_{AA}}_{v_{min}} e^{-v/X}dv \right)^{f^2N} \cdot \non \\
&\cdot& \left( \int^{v_{AB}}_{v_{min}} e^{-v/X}dv
\right)^{2f(1-f)N} \left( \int^{v_{BB}}_{v_{min}} e^{-v/X}dv
\right)^{(1-f)^2N} \label{eq:part_fun_two_species}. \eea Not only
does $Z$ not depend on $z$ when $z$ is assumed to be uniform for
all types of grains, it can easily be seen that when different
values of $z$ are assigned to the different types of grains, the
same expression is obtained. This is an important result in the
analogy between the configurational statistical mechanics of a
granular mixture and the Ising model or a binary alloy. In the
Ising model spins are arranged on an ordered lattice, all spins
have the same number of nearest neighbors, $z$, and the total
energy is determined by the states of the spins on all the lattice
sites. In a granular system, on the other hand, the disordered
spatial configuration determines the volume of the system and
every type of grain may have a different number of nearest
neighbors, or a different coordination number, which is related to
the volume of the Voronoi cell around it, or to the volume it
occupies. Due to the different friction coefficients, $z$ varies
between the different types of grains, as is indicated in
\cite{silbert_2002} where the friction dependence of the average
number of contacts is investigated. The number of contacts of
every type and the contribution of every contact to the total
volume depend inversely on $z$, and since only their product
enters the total volume of the system, the resulting partition
function does not depend on the values of $z$ for the different
types of grains \footnote{An alternative way of deriving the
partition function is to obtain the total volume of the system
from summing over the contributions of all grains to the volume
and not by summing over the contributions of the contacts. In
order to do this we assume all the grains surrounding every grain
are of a single species. Under this assumption and in the
mean-field approximation, the number of A grains surrounded by A
grains is $f^2N$, the number of A grains surrounded by B grains is
$f(1-f)N$, the number of B grains surrounded by A grains is
$f(1-f)N$ and the number of B grains surrounded by B grains is
$(1-f)^2N$. Since the volume of the Voronoi cell around every $i$
grain surrounded by $j$ grains is bounded between $v_{min}$ and
$v_{ij}$, the expression in eq. (\ref{eq:part_fun_two_species})
immediately follows.}. We will soon see where such physical
differences between the granular mixture and the Ising model do
affect the resulting behaviour of the system.

Using the partition function we now evaluate the free energy: \bea
Y(f,X,N) &=& -X \log (Z) = \non\\ &=& XN [f\log(f)+(1-f)\log(1-f)
+ \non\\ &+& 2f(1-f)R(X) - \log(X) - \non\\ &-& fR_{AA}(X) -
(1-f)R_{BB}(X)] \label{eq:free_ener_two_species},\eea where we
have used the notations
$R_{ij}(X)\equiv\log\left(e^{-v_{min}/X}-e^{-v_{ij}/X}\right)$ and
$R(X)\equiv(R_{AA}+R_{BB})/2-R_{AB}$, and the Stirling's
formula has been used to evaluate $\log N!$ for large $N$.
Defining $\Delta v_{ij} \equiv v_{ij} - v_{min}$ we see that \bea
R(X)=\log \left(\frac{\sqrt{\left(1-e^{-\Delta v_{AA}/X}\right)
\left(1-e^{-\Delta v_{BB}/X}\right) }}{\left(1-e^{-\Delta
v_{AB}/X}\right)} \right) \label{eq:R_of_X}. \eea

We would now like to minimize $Y$ for given overall compositions,
$N_{A}$ and $N_{B}$, and for a given compactivity, $X$. Eq.
(\ref{eq:free_ener_two_species}) has been derived under the
mean-field approximation, therefore it describes the free energy
of a region with uniform concentration, $f$. Formally we should
now define the local concentration as a spatially dependent coarse
grained function, $f(\overrightarrow{r})$, and the free energy as
a functional of it, $Y[f(\overrightarrow{r})]=\int
Y(f(\overrightarrow{r}))d\overrightarrow{r}$, and to require that
$f(\overrightarrow{r})$ minimize $Y$ under the constraint that
$\frac{1}{V}\int
f(\overrightarrow{r})d\overrightarrow{r}=N_A/N$. The
spatial integral over all the system of the last three terms in
eq. (\ref{eq:free_ener_two_species}), which are linear in $f$, is
independent of the function $f(\overrightarrow{r})$. Therefore
these terms do not contribute to the minimization of Y, and we
need only consider the contribution of the first three terms to Y.
We thus obtain an expression similar to the mean-field free energy
of an Ising model or of a binary alloy \cite{kubo}. The
``equilibrium'' concentration is now determined from: \bea
\frac{\partial Y}{\partial f} =
XN\left[\log\left(\frac{f}{1-f}\right) +2(1-2f)R(X)\right] = 0
\label{eq:df_df_eq_0}, \eea which is equivalent to: \bea
2f-1=\tanh\left(R(X)(2f-1)\right)\label{eq:tanh}.\eea For $R(X)<1$
the only solution to this equation is $f=0.5$, which corresponds
to mixing, while for $R(X)>1$ two different solutions exist and
the systems segregates into regions with these two minimizing
concentrations. Therefore, the condition for segregation is that
$R(X)>1$, where $R(X)$ is given by eq. (\ref{eq:R_of_X}).

Since $R(X \rightarrow 0)=0$, no segregation occurs in the limit
of low compactivities. Contrast this to the behavior of the Ising
model and binary alloys, where phase separation exists at low
temperatures, and specifically in the limit of zero temperature
\cite{kubo}. Here there is a minimal critical compactivity, above
which segregation occurs, rather than the maximal critical
temperature, bellow which phase separation occurs in the Ising
model. The basic difference between the model presented here for
granular materials and the Ising model is as follows: In the Ising
model the energy of the system is determined only from its
topological state. Namely, once the topology of which element is
the nearest neighbor of which other elements is specified, the
total energy of the system is determined. In the mean-field
approximation these topological states are described by the
concentration, hence the energy depends only on it. In the
granular model presented here a topological state, specifying the
types of contacts around every grain, allows for a range of
volumes for the Voronoi cells around every grain and therefore for
a range of values for the overall volume of the system. As in the
Ising model, the topological state is specified by the
concentration in the mean-field approximation, however the total
volume of the system depends both on the concentration and on the
compactivity, X.

The reason for segregation only at high compactivities may be
understood to stem from the range of possible volumes in the
following way. The Voronoi cells around the grains all have the
same minimal volume, $v_{min}$, determined from geometrical
considerations. The probability for finding a Voronoi cell in an
exited state with a larger volume, $v>v_{min}$, decays as
$e^{-v/X}$. Therefore at the limit $X \rightarrow 0$ all Voronoi
cells are expected to be found in their ground state, namely to
have a volume $v=v_{min}$, independent of the friction of the
grains, which only determines the maximal Voronoi cell volume,
$v_{max}(\mu)$. Hence, at low compactivities the differences
between grains vanish and the two species mix homogeneously.
Segregation may occur only at high compactivities, where exited
states, which exist due to friction, have a greater
``thermodynamic'' weight. This occurrence of segregation at high
compactivities rather than at low compactivities may explain the
frequent appearance of segregation in granular systems, which
typically have significant compactivities, since $X=0$ corresponds
to a crystalline packing.

Since for $R>0$, $R(X)$ is a monotonic function (as can be seen
from eq. (\ref{eq:R_of_X}) which begins at zero, and since the
condition for segregation is that $R>1$, there is only a minimal
critical compactivity for segregation (and no upper bound above
which there is no segregation) so we may determine whether
segregation is possible by investigating the high compactivity
limit of $R(X)$. This limit may easily be evaluated to be equal to
\bea R_{\infty} \equiv R(X\rightarrow\infty)=\log
\left(\frac{\sqrt{\Delta v_{AA}\Delta v_{BB}}}{\Delta v_{AB}}
\right) \label{eq:R_asymp}, \eea and the condition for the
existence of a critical compactivity for segregation for given
values of the friction coefficients is that $R_{\infty}>1$.

This condition may be understood in the following way: The free
energy, $Y=V-XS$, includes a volume ($V$) term and an entropy
($S$) term. At high compactivities the entropy dominates and
minimizing the free energy is equivalent to maximizing the
entropy. The entropy includes two factors, a ``combinatoric
entropy'' related to the topological state of the system, namely
which grain is in contact with which other grains, and a
``geometric'' entropy related to the variety of volumes the
Voronoi cell around every grain may have within one specific
topological state (This is not the case for the Ising model, where
the topological state determines the state of the system, and the
total entropy includes only the combinatoric entropy, or the
entropy of mixing \cite{christian}).

The entropy's dependence on concentration at the limit
$X\rightarrow\infty$ is given by \bea
S&=&-N[f\log(f)+(1-f)\log(1-f)+ \non\\ &+& 2f(1-f)R_{\infty}]
\label{eq:entropy_asymp_two_species}.\eea The first two terms are
the combinatoric entropy or entropy of mixing, while the last term
is the geometric entropy. The geometric entropy plays a similar
role to the role of the energy in the Ising model, as can be seen
in the expression for the free energy in the Ising model
\cite{kubo}: \bea F&=&N[kT(f\log(f)+(1-f)\log(1-f))- \non\\ &-&
2f(1-f)Jz] \label{eq:free_ener_ising},\eea where $k$ is the
Boltzman constant, $T$ is temperature, $J$ is the interaction
energy between neighboring spins and $z$ is the number of nearest
neighbors per site. Ising systems exhibit phase separation when
$kT<-Jz$, which may now be seen to be completely analogous to the
condition $R_{\infty}>1$ for segregation in granular binary
mixtures at high compactivities.


\subsection{Phase Diagram}

\begin{figure}[htb]
\includegraphics[width=8cm]{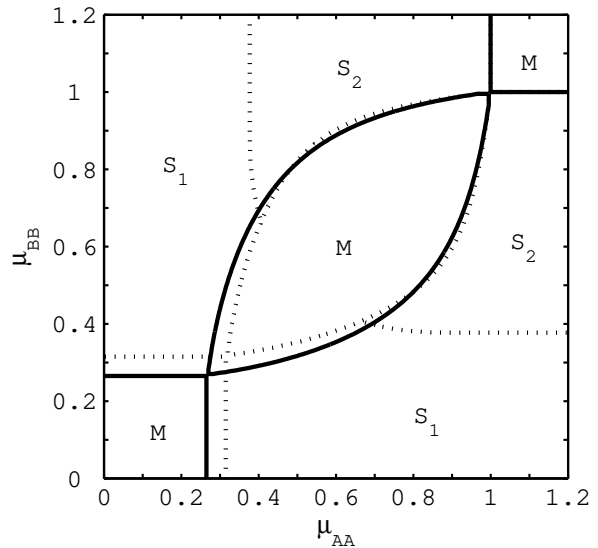}
\caption{\label{fig:mu_aa_mu_bb_plane} Phase diagram of the
$\mu_{AA}-\mu_{BB}$ plane under the assumption that
$\mu_{AB}=min(\mu_{AA},\mu_{BB})$ for 2D (solid lines) and 3D
(dotted lines). M indicates mixing, $S_1$ and $S_2$ indicate
segregation. M regions are always mixed. In 2D S regions are
segregated at high enough compactivity (see text). In 3D $S_1$
regions are always segregated, while $S_2$ regions are segregated
only at high enough compactivity (see section
\ref{sec:eliminate}).}
\end{figure}

We would now like to generate a phase diagram for segregation and
to check for which values of $\mu_{AA}$, $\mu_{BB}$ and $\mu_{AB}$
segregation is possible. From mechanical considerations we would
expect $\mu_{AB}$ to lie between $\mu_{AA}$ and $\mu_{BB}$. Since
$v_{max}(\mu)$ is monotonic, $\Delta v (\mu)$ is monotonic as
well, and $\Delta v_{AB}$ is bounded between $\Delta v_{AA}$ and
$\Delta v_{BB}$. First we notice that if
$\mu_{AB}=max(\mu_{AA},\mu_{BB})$, $R_{ \infty }<0$ and no
segregation is expected. For given values of $\mu_{AA}$ and
$\mu_{BB}$, $R_{ \infty }$ varies monotonicaly with $\mu_{AB}$,
and a necessary condition for segregation in intermediate values
of $\mu_{AB}$ is that there is segregation at
$\mu_{AB}=min(\mu_{AA},\mu_{BB})$. In this case \bea
R_{\infty}=\log \left( \sqrt{\frac{max(\Delta v _{AA},\Delta
v_{BB})}{min(\Delta v_{AA},\Delta v_{BB})}}\right).\eea Fig.
\ref{fig:mu_aa_mu_bb_plane} displays the regions in the
$\mu_{AA}-\mu_{BB}$ plane where $R_{\infty}>1$, and in which
segregation may occur. When $R_{\infty}<1$ the two species are
always expected to be mixed. When $R_{\infty}>1$ segregation
occurs at the limits of $\mu_{AB}=min(\mu_{AA},\mu_{BB})$ and
$X\rightarrow\infty$. Segregation also occurs at larger values of
$\mu_{AB}$ and at finite values of $X$. The existence of
segregation may be verified for the general case by checking
whether the expression for $R(X)$ in eq. (\ref{eq:R_of_X}) is
greater than one. Here we only demonstrate our model results for
the limits described above. As will be shown in the following
subsection, in some cases segregation occurs for every
compactivity. The phase diagram shows that mixtures of grains with
close friction coefficients will mix and what difference in
friction coefficients is needed in order to achieve segregation.

Within the framework of the mechanical model used here, frictional
forces are relevant only for $0.3 \lesssim \mu \leq 1$. If
$\mu_{AA} \lesssim 0.3$ or $\mu_{BB} \lesssim 0.3$ it is
questionable whether this mechanical model is valid. If both
$\mu_{AA} \geq 1$ and $\mu_{BB} \geq 1$ the friction has an
identical effect no matter what the values of $\mu_{AA}$ and
$\mu_{BB}$ are, therefore the two species mix. If only one of
$\mu_{AA}$ or $\mu_{BB}$ is greater than one, the maximal volume
per grain of the corresponding species is unbounded, while for the
second species it is bounded, leading to $R_{\infty}>1$, and hence
to segregation. These regions are designated as well in fig.
\ref{fig:mu_aa_mu_bb_plane}.


\subsection{Eliminating the Compactivity}
\label{sec:eliminate}

Given the mechanical properties of the grains, namely $\mu_{AA}$,
$\mu_{AB}$ and $\mu_{BB}$, we have employed a mean field
approximation in order to solve the Edwards statistical mechanics
description of a granular mixture and to obtain the concentration
(or concentrations in case of segregation) the system will be
found at as a function of compactivity. Even though there is
evidence that it may be controlled experimentally
\cite{nowak_1997,edwards_grinev_PRE_1998}, the compactivity is
still not a measurable quantity, and we would like to reach a
description of segregation independent of the notion of
compactivity.

For a single species we have seen that the total volume of the
system, and correspondingly its volume fraction, depend on
compactivity. We shall now derive the compactivity dependence of
the volume fraction in a binary mixture, and use it in order to
eliminate the compactivity from the description of segregation.
This will result in a relation between concentrations and volume
fraction, which may be measured experimentally.

The average volume per grain may easily be derived from the
partition function (eq. (\ref{eq:part_fun_two_species})) as: \bea
<v> &=& \frac {<V>}{N} = \frac {1}{N}
X^2 \frac {\partial\ln(Z)}{\partial X} = \non \\
&=& X+f^2Q_{AA}(X)+2f(1-f)Q_{AB}(X)+ \non\\ &+& (1-f)^2Q_{BB}(X)
\label{eq:avrg_vol_two_specie}, \eea where \bea Q_{ij}(X) &\equiv&
X^2\frac{\partial R_{ij}(X)}{\partial X}= \non\\ &=&
\frac{v_{min}e^{-v_{min}/X}-v_{ij}e^{-v_{ij}/X}}
{e^{-v_{min}/X}-e^{-v_{ij}/X}}\label{eq:Q_of_X}.\eea As for a
single species, when $X \rightarrow 0$, $Q_{ij}(X) \rightarrow
v_{min}$ and $<v> \rightarrow v_{min}$. Moreover, when $X
\rightarrow \infty$, $Q_{ij}(X) \approx v_{ij}+v_{min}-X$, hence
$<v> \rightarrow v_{min}+f^2v_{AA}+2f(1-f)v_{AB}+(1-f)^2v_{BB}$.

\begin{figure}[htb]
\includegraphics[width=8cm]{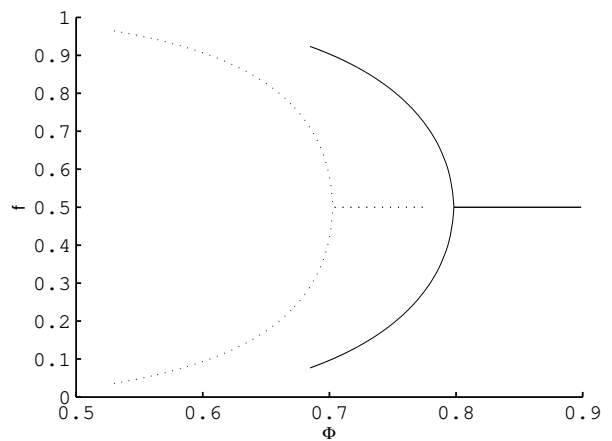}
\caption{\label{fig:seg_vs_vol} Concentration, $f$, vs. volume
fraction, $\Phi$, for $\mu_{AA}=0.35$, $\mu_{BB}=0.75$,
$\mu_{AB}=0.35$ for 2D (solid lines) and 3D (dotted lines). For
low compactivities (or high volume fractions) a single
concentration is possible and the two species mix. For high
compactivities (or low volume fractions) two concentrations are
possible and segregation occurs.}
\end{figure}

The concentration $f$ may be expressed as a function of $X$ by
using eq. (\ref{eq:tanh}). Eliminating $X$ between eq.
(\ref{eq:tanh}) and (\ref{eq:avrg_vol_two_specie}), we obtain $f$
as a function of the volume fraction, $\Phi$. This is plotted in
fig. \ref{fig:seg_vs_vol} for typical values of the friction
coefficients ($\mu_{AA}=0.35$, $\mu_{BB}=0.75$ and
$\mu_{AB}=0.35$). At low compactivities the volume fraction is
high and the system is mixed. As the compactivity is raised the
volume fraction is reduced and at some critical point segregation
commences, and two different concentrations appear. These two
values are the concentrations in different domains of the system.
As the compactivity is raised above the critical point the two
concentrations separate one from each other, enhancing the
difference between the different domains. We identify the critical
compactivity for segregation, $X_c$, above which segregation
occurs and the critical volume fraction for segregation, $\Phi_c
\equiv \Phi(X_c)$, below which segregation occurs. An important
point to note in the case demonstrated in fig.
\ref{fig:seg_vs_vol} is that in 3D $\Phi_c \simeq 0.7$, which is
larger than the volume fraction of 3D RCP, $\Phi_{RCP} \simeq
0.64$. Granular materials typically do not reach volume fractions
higher than that of RCP, therefore in this case the compactivity
is always high enough so that the volume fraction is smaller than
$\Phi_c$ and segregation always occurs. We now use this argument
in order to divide the $\mu_{AA}-\mu_{BB}$ plane into regions
where $\Phi_c>\Phi_{RCP}$ (denoted $S_1$ in fig.
\ref{fig:mu_aa_mu_bb_plane}), which are segregated for all $X$,
and regions where $\Phi_c<\Phi_{RCP}$ (denoted $S_2$ in fig.
\ref{fig:mu_aa_mu_bb_plane}), which are segregated only above some
$X_c$ (or only under some $\Phi_c$). Such considerations may not
be used in 2D, where full crystalization is achieved
experimentally.

\section{Conclusions:}
\label{sec:conc}

We have used Edwards' statistical mechanics hypothesis together
with a simple mechanical model to describe the role of friction in
2D and 3D granular materials. For a single species this describes
the decrease in volume fraction with increasing friction
coefficient and with increasing compactivity. An experimental test
of the ideas presented is most easily interpreted for systems
differing in friction coefficient but with the same compactivity.
It is intriguing to consider whether identical preparation would
lead to equal compactivities for systems differing only in
friction coefficients.

In addition, the model has been used in order to investigate
segregation in binary mixtures of grains differing in frictional
properties. Unlike mixtures of grains differing in size, which may
be mapped to the Ising model, frictional differences between
grains result in larger entropy for the rougher grains; therefore,
these systems may not be mapped exactly onto the Ising model, we
find that segregation occurs above a critical compactivity and not
below it. A phase diagram for segregation vs. friction
coefficients of the two species, which may be tested
experimentally, has been generated. By eliminating the
compactivity, we have also provided a relation between the volume
fraction and the nature of mixing or segregation. This relation
both provides an option for experimental validation of the model
and of the statistical hypothesis and allows to identify mixtures
which are expected to segregate at every compactivity.

The geometrical and mechanical models used to describe friction in
this paper are more qualitative than quantitative, and more
sophisticated models may be suggested. However, the qualitative
results obtained here do not depend on their details but only on
basic properties which all such models should have: the minimal
volume does not depend on friction and the maximal volume and the
number of possible states increase with friction. The experimental
validation or invalidation of the aforementioned results may shed
light on the validity of the statistical mechanics proposal of
Edwards.


\begin{acknowledgments}
We would like to thank Yael Roichman, Guy Bunin, Sam Edwards and
Dmitri Grinev for helpful discussions. DL acknowledges support
from grant no. 9900235 of the US-Israel Binational Science
Foundation, grant no. 88/02 of the Israel Science Foundation and
the Fund for the Promotion of Research at the Technion.
\end{acknowledgments}


\appendix
\section{Voronoi Cell Volumes}
\label{sec:geom_calc}

This appendix describes the calculations leading from the
mechanical models in 2D and in 3D to the volume of the
corresponding Voronoi cells around the grains, given in eq.
(\ref{eq:vol_vs_theta}). In 2D the Voronoi cell around a grain is
formed by the tangents to it at the contact points with the
surrounding grains. The segment of the Voronoi cell lying between
two contacts is the shaded rhombus in fig.
\ref{fig:basic_balls_2d}, which has an area of $r^2 \tan \theta$,
where $r$ is the grain radius and $\theta$ is half the angle
between two adjacent contacts. If all angles between adjacent
contacts are equal to $2\theta$, the Voronoi cell is comprised of
$\pi / \theta$ such segments. Even though this is an integer
number only for discrete values of $\theta$, we use this for
continuous values of $\theta$. The resulting area of the Voronoi
cell is $v_{vor}^{2D}=\frac {\pi r^2 \tan \theta}{\theta}$.

\begin{figure}[htb]
\includegraphics[width=8cm]{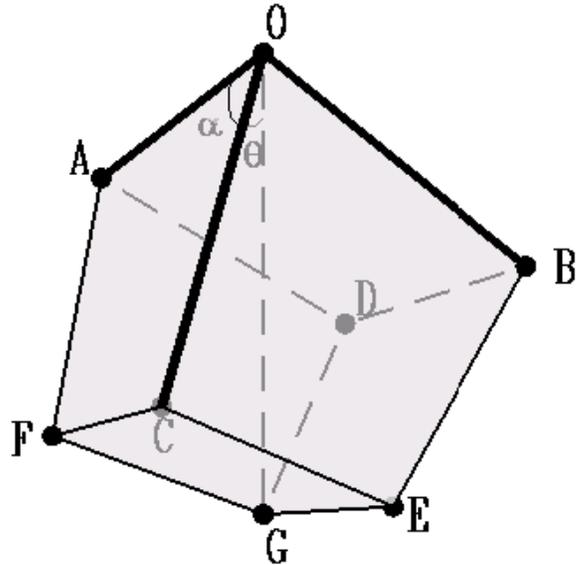}
\caption{\label{fig:vor_cell_3d} The segment of the 3D Voronoi
cell around the grain centered at $O$ and blocked between the
three contacts $A$, $B$ and $C$. The tangent planes at the contact
points meet the planes formed by the directions to the contact
points at $D$, $E$ and $F$ and all meet at $G$.}
\end{figure}

In 3D we use a triangulation of the contact points on the surface
of every grain, and in analogy to the 2D case, we consider the
segment of the Voronoi cell bounded between three contacts (see
fig. \ref{fig:vor_cell_3d}).
We assume the three contact points $A$, $B$ and $C$ form an
equilateral triangle, therefore $\angle AOG = \angle BOG = \angle
COG = \theta$, where $O$ is the center of the grain and $G$ lies
on the line directed from $O$ to the center of the triangle
$\vartriangle ABC$. The segment of the Voronoi cell is the region
bounded between the three directions $OA$, $OB$ and $OC$ (or the
planes $AOB$, $BOC$ and $COA$) and the tangent planes $ADGF$,
$BEGD$ and $CFGE$ (which due to symmetry meet at the point $G$,
which is located along the direction from $O$ to the center of the
triangle $\vartriangle ABC$). The segment of the Voronoi cell may
be divided into six tetrahedra of equal volume all having the
common edge $OG$: $OADG$, $OAFG$, $OBDG$, $OBEG$, $OCEG$ and
$OCFG$. Its volume is hence given by \bea
v_{seg}&=&6v_{OCEG}=6\frac{S_{OCG}H_E}{3}= \non\\ &=& r^2 \tan
\theta \cdot \frac{\sqrt{3}r\sin\theta}{1+3\cos^2\theta}
\label{eq:seg_vol}, \eea where $S_{OCG}$ is the area of the
triangle $\vartriangle OCG$ and $H_E$ is the distance of $E$ from
the plane of $\vartriangle OCG$. In order to calculate the total
volume of the Voronoi cell we calculate the number of segments it
is comprised of through the solid angle every such segment
occupies. The angle $\angle AOB = \angle BOC = \angle COA =
\alpha$ may be calculated from the scalar product of any two of
the vectors $OA$, $OB$ and $OC$ and is given by
$\cos\alpha=(3\cos2\theta+1)/4$. The solid angle formed by
these three vectors is $\Omega = 4 \tan^{-1}
\left(\sqrt{\tan(3\alpha/4)\tan^3(\alpha/4)}\right)$.
Therefore the Voronoi cell volume is \bea
v_{vor}^{3D}&=&v_{seg}\frac{4\pi}{\Omega}= \non\\ &=& \frac{\pi
r^3 \sqrt{3}\sin\theta\tan \theta}{(1+3\cos^2\theta)\tan^{-1}
\left(\sqrt{\tan\frac{3\alpha}{4}\tan^3\frac{\alpha}{4}}\right)}
\label{eq:vor_vol}. \eea


\bibliography{friction}
\bibliographystyle{prsty}

\end{document}